# Plasmonic Superradiance of Two Emitters Near Metal Nanorod

*Igor Protsenko*[*,1,2,3], *Shunping Zhang* [3], *Alexander Uskov*[1,4], *Xuewen Chen*[5], *and Hongxing Xu*[3].

[1]Lebedev Physical Institute, Leninsky prospect, 53, Moscow, 119991, Russia

[2]Advanced Energy Technologies ltd, Cheremuskinsky pass 5, Moscow, 117036, Russia

[3] School of physics and technology, Wuhan University, Wuhan 430072, China,

[4]ITMO University, Kronverksky pr. 49, St. Petersburg, 197101, Russia,

[5]School of Physics, Huazhong University of Science and Technology, 430074, Wuhan, China

[*]*Email address:protsenk@sci.lebedev.ru*

Quantum emitters, such as q-dots and dye molecules, in the immediate vicinity of plasmonic nanostructures, resonantly excite surface plasmon-polaritons (SPPs) under incoherent pump. The efficiency in the excitation of SPPs increases as the number of the emitters, because the SPP field synchronizes emission of the coupled emitters, in analogy with the superadiance (SR) of coupled emitters in free space. Using fully quantum mechanical model for two emitters coupled to a single gold nanorod, we predict up to 15% increase in the emission yield of single emitter compared to only one emitter coupled to the nanorod due to plasmonic SR. (XW: I use emission yield because the quantum efficiency is one for emitters in free space and there is no room for enhancement). Such emission enhancement is stationary and should be observable even with strong dissipation and dephasing under incoherent pump. Solid-state quantum emitters with blinking behaviors may be utilized to demonstrate such plasmonic SR emission enhancement. Plasmonic SR may find implications in the excitation of non-radiative modes in plasmonic waveguides; and lowing threshold of plasmonic nanolasers.

1. **Introduction**

Superradiance (SR), or collective spontaneous emission of photons takes place at transitions between "bright" (Dicke) states of encemble of $N$ coupled dipole emitters, confined and regularly distributed in a volume small compared to the wavelength of the emission. Emitters interact with each other through electromagnetic field by the dipole-dipole interaction [1-3]. Transitions between collective "bright" states of dipoles correspond to coherent oscillations of their dipole moments resulting in a larger total dipole moment.. This leads to fast radiative decay and forms intenstive SR pulse [4]. One may suggest placing the coupled dipole emitters in a microcavity, excite them continuously and utilize the large collective dipole moment for increasing the efficiency of the stationary light emission. Modern technologies allow to create large varieties of microcavities with active media and thus it motivates us to investigate the SR in mesoscopic systems [5], in micro- and nanocavities [6-12].

One particular interesting type of cavity is the so-called "plasmonic" nanocavity or waveguide: metal nanostructure with plasmon-polariton resonance. It has nano-scaled size in one or all three dimentions and can interact resonantly with emitters through the strong near field. Plasmonic cavities and waveguides have been proposed as buliding blocks for highly integrated photonic circuits [13]. However, due to absorption in metal, plasmonic cavities have high losses and low quality factors. This hinders some applications of plamonic nanocavities but in the context of the current study provides proper conditions for the realization of SR [6-8] which, in turn, may help to overcome the loss problem at some extent. Indeed, a low-Q cavity provides photon bath, similar with free space bath. Bath degrees of freedom can be eliminated adiabatically, which means the same photon never comes back to emitter and therefore the emitters decay radiatively [14-16]. Cavity modes at resonance have higher density of states [17] and lead, in particular, to enhanced spontaneous emission of single resonant emitter which can also be understood as SR [18]. Collective spontaneous emission of several emitters can also be enhanced near metal nanoparticle [19, 20].

Different from SR in vacuum, plasmonic SR is allways affected by dissipative environment due to the absorption of radiation in metal, reducing the SR efficiency [19,20].

SRs near metal nanoparticle, as well as in vacuum, are usually considered as radiative decay of emitters after initial instant excitation [19]. However, the case most interesting for applications, is SR at stationary regime, where the system is under continuous incoherent pump.

The key signature of SR in vacuum is the increase of radiative decay rate due to the increase of total dipole moment of coupled emitters [1]. For SR in dissipative environment this means the increase of quantum efficiency of radiation per emitter [22]. The decay rate of photoluminescence as a function of the number of interacting QDs (without nearby metal nanostructure) and their respective separation has been investigated in Ref. [23]. However such proof-of principle experiments are difficult because the theory of SR of a few quantum emitters in free space predicts only minor influence to radiative lifetime [3] and the influence becomes even smaller at the prescence of non-radiative decay [22] and reduces further by inhomogeneous broadening and other features for solid-state emitters [24]. High-precision experiments with an isolated atom pair in free space [25] and with superconductive q-bits in 1-D cavity [10] only revealed a marginal change of cooperative lifetime. Thus the increase of density of photonic states near metal nanostructures may be quite helpful to increase the SR efficiency, which is necessary for its experimental observation and its applications.

The purpose of this paper is to investigate stationary SR of two emitters, resonantly coupled, via near field with the localized SPP mode of a nearby metal nanorod. We calculate and show the increase of efficiency of SPP generation per emitter with respect to single emitter near the nanorod. The enhancement is much larger than for SR in free space. Here we consider *plasmonic* SR, i.e., emitters generate plasmons, as in [13], not photons. Our approach can be easily generalized to many emitters and used for calculating the statistical properties of SR as in [6]. We suppose, that each emitter is incoherently pumped (by external field, injected current, etc.) and the frequency of the transition coinsides with the SPP frequency. As an example, we consider quadrupole SPP mode with analytical approximation described in the Appendix. Quadrupole SPP mode is chosen because of its relatively low radiation losses and, therefire, higher Q-factor than the dipole SPP mode. The excitation of higher-order SPP modes can be considered by the same procedure as in the Appendix. For calculations we use simple, but fully quantum-mechanical model, taking into account of all the quantum correlations at the prescence of dissipation. The approximation we apply is the adiabatic elimination of SPP, which is appropriate at weak coupling of emitters with low-Q SPP. Adiabatic elimination of mode in the low-Q cavity is quite usual assumption at the modelling of SR as, for example, in [8].

Fluorescence enhancement of many emitters near metal nanoparticle, respectively to fluorescance in free space, has been calculated in [26], but in the frame of rate equations, without taking into account coherent emitter-emitter coupling through SPP mode. In the meanwhile, coherent phenomena in quantum emitters near metal nanoparticle can be important, as it is shown below for SR and in other examples, as in [27]. Coherent coupling of two emitters with metal nanowire was studied in [17]. There cooperative surface plasmon emission, achieved for the highest β–factor has been mentioned, without, however, solving the complete quantum-mechanical problem for two emitters and without the comparison of quantum efficiency of SPP generation with the case of single emitter coupled to the nanowire.

We determine conditions, when quantum efficiency of SPP generation by two emitters near single metal nanorod is higher, than for two nanorodes with single emitter near each one, at the same emitter-nanorod coupling. Such increase in efficiency is caused by collective radiation of emitters into SPP mode. As numerical parameter we use quantum efficiency of generation of SPP normalized to the number of emitters, this parameter is named below as a "Single emitter Quantum Efficiency" (SQE). We'll find conditions when SQE is larger for two emitters near metal nanoparticle, than for single emitter at the same conditions. We'll see, that the increase in SQE is due to synchronization of chaotic dipole oscillations of emitters interacting with each other through the near field of SPP. This is very similar to how two dipoles, close to each other in free space, became self-synchronized by the dipole-dipole interaction, and produce SR of photons. However here we study SR of SPPs. When the SPP radiates into free space, we observe SR of photons. We take into account of the dephasing , non-radiative decay at the transitions of emitters and the absorption in metal. It turns out that necessary condition for the stationary SR is the population inversion at transitions of emitters, which makes SR in our case similar with lasing.

We describe SPP in terms of "plasmons": quanta of oscillations of full dipole moment of metal nanoparticle. Electric field indiced by oscillations of plasmons can be recovered outside (and, if necessary, inside) the nanoparticle.

In Section 2 we present the system Hamiltonian and derive Heisenberg operator equations of motion for $N$ emitters near a metal nanorod. We adiabatically eliminate SPP operator at the assumption of strong dissipation and weak coupling. The case of strong coupling of SPP and emitterts is considered in [28]. Taking Heisenberg operator equations for $N$ emitters from Section 2, we derive equations for expectation values of single- and binary products of operators of two emitters in Section 3. Section 4 contains main results of the paper: conditions of increase of quantum efficiency per emitter due to SR. There we find the stationary solution of equations for two emitters when emitters are equally coupled with SPP, leaving extended case of emitters non-equally coupled with SPP for the future. Discussion of results and the conclusion are given in Sections 5 and 6, respectively.

In the Appendix we present analytical description of the quadrupole SPPs of metal nanorod , where was investigated numerically in Ref. [29].

## 2. Hamiltonian and Equations of motion for N emitters

Hamiltonian of interaction of SPP of the nanorod with $N$ two-level emitters, written in rotating wave approximation, taking SPP frequency to be a carrier frequency, is

$$H_{RWA} = \hbar \sum_{i=1}^{N} \left[ -\delta_i n_i^a - \Omega_i \left( \sigma_i^+ a + c.c. \right) \right] + \hat{\Gamma}_{LPPM} + \sum_{i=1}^{N} \hat{\Gamma}_{ei} . \qquad (1)$$

Here $n_i^a$ is the operator of population of excited state, $\sigma_i$ is the operator of transition from the excited state $|a\rangle_i$ to the ground state $|b\rangle_i$ of i-th emitter, $a$ is Bose annihilation operator of plasmon, $\delta_i = \omega - \omega_{ei} \ll \omega$ is the detuning between frequency $\omega$ of SPP and emitter transition frequency $\omega_{ei}$ of transition $|a\rangle_i \rightarrow |b\rangle_i$; coupling coefficient $\Omega_i$ (Rabi frequency) is assumed to be a real number. For particular case of SPP – quadrupole mode of the nanorod $\Omega_i = \dfrac{d_e |\vec{E}_i|}{2\sqrt{2}\hbar}$ is found in Appendix, $d_e$ is a dipole matrix element of emitter's transition, $\vec{E}_i$ is the amplitude of electric field of SPP at emitter $i$; $\hat{\Gamma}_{LPPM}$ describes incoherent decay of SPP, $\hat{\Gamma}_{ei}$ describes incoherent decay and pump in $i$-th emitter.

We write Heisenberg operator equations of motion $i\hbar \dot{\hat{A}} = \left[ \hat{A}, H_{RWA} \right]$, where $\hat{A}$ stands for $a$, $\sigma_i$, $\sigma_i^+$, $n_i^a$ and for the operator $n_i^b$ of the population of the low state of $i$-th emitter. We use well-known commutation relations

$$\left[ a, a^+ \right] = 1 ; \quad \left[ \sigma_i, \sigma_i^+ \right] = n_i^b - n_i^a ; \quad \left[ \sigma_i, n_i^b \right] = -\sigma_i ; \quad \left[ \sigma_i, n_i^a \right] = \sigma_i , \qquad (2)$$

and obtain equations of motion:

$$\dot{a} = -\kappa a + i\sum_{i=1}^{N}\Omega_i \sigma_i + F_a$$

$$\dot{\sigma}_i = (i\delta_i - \Gamma)\sigma_i - i\Omega_i(n_i^a - n_i^b)a + F_{i\sigma}$$

$$\dot{n}_i^a = -i\Omega_i(a^+\sigma_i - \sigma_i^+ a) - \frac{1}{\tau_a}n_i^a + \Gamma_p(1-n_i^a) + F_{in_a} \qquad (3)$$

$$\dot{n}_i^b = i\Omega_i(a^+\sigma_i - \sigma_i^+ a) - \frac{1}{\tau_b}n_i^b + F_{in_b}$$

Here $\kappa$ is SPP decay rate, $\Gamma$ is polarization relaxation rate of emitter (the half of the emitter transition linewidth); $\tau_{a,b}^{-1}$ is the relaxation rate of the upper and lower states of emitters, $\Gamma_p$ is the pump rate, the term $\Gamma_p(1-n_i^a)$ takes into account Coulomb blockade at the pump, the same term used, for example, in;[30] $F_{i\beta}$ is Langevin force corresponding to variable $\beta = a, \sigma_i, ...$

Here we suppose the same pump-decay scheme of single emitter as in.[31] Two-level emitter with states $|a\rangle$ and $|b\rangle$ interacts with pump and decay reservoirs with rates $\Gamma_p$, $\tau_{a,b}^{-1}$ shown in **Figure 1**. For simplicity, we suppose that decays to reservoirs are much faster than decays from $|a\rangle_i$ to $|b\rangle_i$, and neglect such decays; relaxation rate of emitter transition dipole moment is

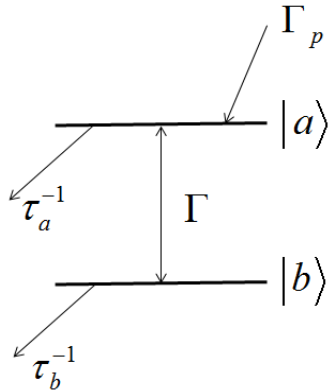

**Figure 1** Scheme of the relaxation and the pump of single emitter.

$$\Gamma = \frac{1}{2\tau_a} + \frac{1}{2\tau_b} + \Gamma_{deph}, \qquad (4)$$

where $\Gamma_{deph}$ is dephasing rate.

It is convenient to introduce population inversion $\Delta_i = n_i^a - n_i^b$ and $s_i = n_i^a + n_i^b$ instead of $n_i^a$ and $n_i^b$. Equation for $s_i$ follows from Equation (3):

$$\dot{s}_i = -\frac{1}{\tau_b}n_i^b - \frac{1}{\tau_a}n_i^a + \Gamma_p\left(1-n_i^a\right) + F_{in_a} + F_{in_b}. \tag{5}$$

Here we are looking for the stationary solutions for expectation values. Then we can set $\dot{s}_i = 0$. As we'll see, $n_i^{a,b}$ and $\Delta_i$ appear in the mean values of products only for commuting operators with $j \neq i$. Langevin forces do not contribute into equations for such products, and we can drop Langevin forces in Equation (5). Than from the stationary Equation (5) we find:

$$n_i^b = \Gamma_p \tau_b - \left(\frac{\tau_b}{\tau_a} + \Gamma_p \tau_b\right) n_i^a \tag{6}$$

and

$$n_i^a = \frac{\Gamma_p \tau_b + \Delta_i}{1 + \tau_b/\tau_a + \Gamma_p \tau_b}. \tag{7}$$

Using Equation (3) we can write:

$$\begin{aligned}
\dot{a} &= -\kappa a + i\sum_{i=1}^{N}\Omega_i \sigma_i + F_a \\
\dot{\sigma}_i &= (i\delta_i - \Gamma)\sigma_i - i\Omega_i \Delta_i a + F_{\sigma_i} \\
0 &= 2i\Omega_i\left(\langle\sigma_i^+ a\rangle - \langle a^+ \sigma_i\rangle\right) - \frac{1}{\tau'}\left(\langle\Delta_i\rangle - \Delta_0\right)
\end{aligned} \tag{8}$$

Here and below $\langle...\rangle$ means expectation value;

$$\Delta_0 = \frac{\Gamma_p \tau_a}{\Gamma_p \tau_a + 1}, \quad \frac{1}{\tau'} = \frac{2}{\tau_a}\frac{1+\Gamma_p \tau_a}{1+\tau_b/\tau_a + \Gamma_p \tau_b}. \tag{9}$$

Supposing $\kappa \gg \Gamma, 1/\tau'$ and weak coupling $\kappa \gg \Omega_i$ we can adiabatically eliminate $a$ from Eqs.(8) by setting $\dot{a} = 0$, so that

$$a = \frac{i}{\kappa}\sum_{i=1}^{N} \Omega_i \sigma_i + \frac{F_a}{\kappa}. \tag{10}$$

We can drop $F_a$ in Equation (10) because of $\langle a^+ a \rangle$ and other expectation values contains $\langle F_{a^+}(t) F_a(t') \rangle = 0$ (assuming zero temperature of plasmonic reservoir). Any other mean value of products with $F_a$ or $F_{a^+}$ are zero, because of operators of emitters commute with $a$ and $a^+$. Thus, instead Equation (10) we can write

$$a = \frac{i}{\kappa}\sum_{i=1}^{N} \Omega_i \sigma_i. \tag{11}$$

Inserting $a$ from Equation (11) into the second and the third ones of Equation (8) and using operator relations $\Delta_i \sigma_i = -\sigma_i$ and $\sigma_i^+ \sigma_i = n_{ai}$ we obtain

$$\begin{aligned}
\dot{\sigma}_i &= \left[i\delta_i - (\Gamma + \gamma_i)\right]\sigma_i + \sum_{j \neq i}\Omega_{ij}\Delta_i\sigma_j + F_{\sigma_i} \\
0 &= -4\gamma_i \langle n_i^a \rangle - \sum_{j \neq i}\Omega_{ij}\left(\langle \sigma_i^+ \sigma_j \rangle + \langle \sigma_j^+ \sigma_i \rangle\right) - \frac{1}{\tau'}\left(\langle \Delta_i \rangle - \Delta_0\right)
\end{aligned}, \tag{12}$$

where we represent coupling constants as

$$\Omega_{ij} = \frac{\Omega_i \Omega_j}{\kappa} = \Omega_{ji} \equiv \sqrt{\gamma_i \gamma_j}. \tag{13}$$

The term $\sim \gamma_i = \frac{\Omega_i^2}{\kappa}$ in the right hand side of the second one of Equation (12) describes spontaneous emission of emitters into SPP. From Equation (13) we see, that the emitter-emitter coupling coefficient is expressed through $\gamma_i$.

Using Equation (11) we write for the mean number of plasmons:

$$n \equiv \langle a^+ a \rangle = \frac{1}{\kappa}\left( \sum_{i=1}^{N} \gamma_i \langle n_i^a \rangle + \sum_{i \neq j} \sqrt{\gamma_i \gamma_j} \langle \sigma_i^+ \sigma_j \rangle \right). \quad (14)$$

Here the term $\sum_i \gamma_i \langle n_i^a \rangle \equiv \sum_i \gamma_i \langle \sigma_i^+ \sigma_i \rangle$ is the total spontaneous emission rate into SPP from all emitters radiated individually. The interference term $\sum_{i \neq j} \sqrt{\gamma_i \gamma_j} \langle \sigma_i^+ \sigma_j \rangle$ is a contribution of coherent collective emission. For uncorrelated chaotic oscillations, the term $\sum_{i \neq j} \sqrt{\gamma_i \gamma_j} \langle \sigma_i^+ \sigma_j \rangle = 0$ and then the SQE of two emitters is the same as SQE of single emitter near the nanorod.

From Equation (14) and the second one of Equation (12) we find the number of generated plasmons averaged by the number of emitters (N) $n^{(N)} \equiv n / N$:

$$n^{(N)} = \frac{1}{4\kappa\tau'}\left( \Delta_0 - \Delta^{(N)} \right) > 0, \quad (15)$$

where

$$\Delta^{(N)} = \frac{1}{N}\sum_{i=1}^{N} \langle \Delta_i \rangle. \quad (16)$$

From Equation (15) and (16) we obtain simple formula for *Relative Quantum Efficiency* (RQE) per one of $N > 1$ emitters: relatively to single emitter near the nanorod

$$R \equiv \frac{n^{(N)}}{n^{(1)}} = \frac{\Delta_0 - \Delta^{(N)}}{\Delta_0 - \Delta^{(1)}}. \quad (17)$$

3. **Equations of motion for expectation values of two emitters**

From here we restrict ourselves to only two emitters near the nanorod, i.e., $N = 2$. Let us find the stationary number of plasmons $n^{(2)}$ per emitter. According to Equation (15), in order to calculate $n^{(2)}$ we have to find $\Delta^{(2)} = 0.5\sum_{i=1}^{2}\langle\Delta_i\rangle$ from Equation (12), so that we need equation for

$$\Sigma \equiv \langle\sigma_1^+\sigma_2\rangle + \langle\sigma_2^+\sigma_1\rangle. \tag{18}$$

Using the first one of Equation (12) and differentiating the product $\sigma_1^+\sigma_2$ we obtain

$$\frac{d}{dt}\langle\sigma_1^+\sigma_2\rangle = -(2\Gamma+\gamma_1+\gamma_2)\langle\sigma_1^+\sigma_2\rangle + \sqrt{\gamma_1\gamma_2}\left(\langle\Delta_1\sigma_2^+\sigma_2\rangle + \langle\Delta_2\sigma_1^+\sigma_1\rangle\right). \tag{19}$$

From here we suppose, for simplicity, $\delta_i \ll \max\{\Gamma,\gamma_i\}$ and set $\delta_i = 0$.

In equations for expectation values of operator products for different emitters, like Equation (19), the contribution from Langevin forces disappears, because of the operators of different emitters commute.

In Equation (19) we replace $\langle\sigma_i^+\sigma_i\rangle = \langle n_i^a\rangle$, take $\langle n_i^a\rangle$ from Eq.(7) and obtain

$$\frac{d\Sigma}{dt} = -(2\Gamma+\gamma_1+\gamma_2)\Sigma + 4\frac{\sqrt{\gamma_1\gamma_2}}{\theta}\left[\langle\Delta_1\Delta_2\rangle + \Gamma_p\tau_b\frac{\langle\Delta_1\rangle+\langle\Delta_2\rangle}{2}\right], \tag{20}$$

where

$$\theta = 1 + \tau_b/\tau_a + \Gamma_p\tau_b. \tag{21}$$

From Equation (20) we see, that we need equation for $\langle\Delta_1\Delta_2\rangle$. We take the second one of Equation (12), write it for operators, differentiate operator products and find

$$\frac{d}{dt}\langle \Delta_1 \Delta_2 \rangle = \left\langle \left[ -4\gamma_1 n_{a1} - 2\sqrt{\gamma_1\gamma_2}\left(\sigma_1^+\sigma_2 + \sigma_2^+\sigma_1\right) - \frac{1}{\tau'}(\Delta_1 - \Delta_0) \right]\Delta_2 \right\rangle +$$
$$\left\langle \Delta_1 \left[ -4\gamma_2 n_{a2} - 2\sqrt{\gamma_1\gamma_2}\left(\sigma_1^+\sigma_2 + \sigma_2^+\sigma_1\right) - \frac{1}{\tau'}(\Delta_2 - \Delta_0) \right] \right\rangle$$

We replace in above expression $\sigma_i \Delta_i = \sigma_i$, $\sigma_i^+ \Delta_i = -\sigma_i^+$, $\Delta_i \sigma_i^+ = \sigma_i^+$, $\Delta_i \sigma_i = -\sigma_i$ so that

$$\frac{d}{dt}\langle \Delta_1 \Delta_2 \rangle = \left\langle \left[ -4\gamma_1 n_1^a - \frac{1}{\tau'}(\Delta_1 - \Delta_0) \right]\Delta_2 \right\rangle - 4\sqrt{\gamma_1\gamma_2}\left\langle \sigma_1^+\sigma_2 - \sigma_2^+\sigma_1 \right\rangle + \left\langle \Delta_1 \left[ -4\gamma_2 n_2^a - \frac{1}{\tau'}(\Delta_2 - \Delta_0) \right] \right\rangle$$

One can see from Equation (19), that the stationary $\langle \sigma_1^+\sigma_2 - \sigma_2^+\sigma_1 \rangle = 0$, therefore

$$\frac{d}{dt}\langle \Delta_1 \Delta_2 \rangle = -4\left[ \frac{1}{\theta}(\gamma_1 + \gamma_2) + \frac{1}{2\tau'} \right]\langle \Delta_1 \Delta_2 \rangle - \frac{4\Gamma_p \tau_b}{\theta}\left( \gamma_1 \langle \Delta_2 \rangle + \gamma_2 \langle \Delta_1 \rangle \right) + \frac{\Delta_0}{\tau'}\left( \langle \Delta_2 \rangle + \langle \Delta_1 \rangle \right),$$

where we expressed $n_i^a$ through $\Delta_i$ by Equation (7). Thus we obtain closed set of equations for mean values of single operators and for binary products of them:

$$\frac{d\Sigma}{dt} = -(2\Gamma + \gamma_1 + \gamma_2)\Sigma + \frac{4\sqrt{\gamma_1\gamma_2}}{\theta}\left[ \langle \Delta_1 \Delta_2 \rangle + \Gamma_p \tau_b \frac{\langle \Delta_1 \rangle + \langle \Delta_2 \rangle}{2} \right]$$
$$\frac{d}{dt}\langle \Delta_1 \Delta_2 \rangle = -4\left( \frac{\gamma_1 + \gamma_2}{\theta} + \frac{1}{2\tau'} \right)\langle \Delta_1 \Delta_2 \rangle - \frac{4\Gamma_p \tau_b}{\theta}\left( \gamma_1 \langle \Delta_2 \rangle + \gamma_2 \langle \Delta_1 \rangle \right) + \frac{\Delta_0}{\tau'}\left( \langle \Delta_2 \rangle + \langle \Delta_1 \rangle \right) \quad (22)$$
$$\langle \dot{\Delta}_1 \rangle = -\frac{4\gamma_1}{\theta}\left( \langle \Delta_1 \rangle + \Gamma_p \tau_b \right) - 2\sqrt{\gamma_1\gamma_2}\Sigma - \frac{1}{\tau'}\left( \langle \Delta_1 \rangle - \Delta_0 \right)$$
$$\langle \dot{\Delta}_2 \rangle = -\frac{4\gamma_2}{\theta}\left( \langle \Delta_2 \rangle + \Gamma_p \tau_b \right) - 2\sqrt{\gamma_1\gamma_2}\Sigma - \frac{1}{\tau'}\left( \langle \Delta_2 \rangle - \Delta_0 \right)$$

In the next Section we'll find and investigate the stationary solution of Equation (22). Considering similar problem for N emitters we have to write equations for mean values of up to N-operator products. Complexity of such equations grows quickly with N. One can expect, however, that the contribution of high-order correlations become less important for high N, so that they can be truncated as in [5].

## 4. The stationary solution and RQE for symmetric positions of emitters

Here we consider symmetric case, when emitters have the same interaction with SPP electric field, so that $\gamma_1 = \gamma_2 = \gamma$ and $\langle \Delta_1 \rangle = \langle \Delta_2 \rangle \equiv \Delta^{(2)}$. Now Equation (22) read

$$\frac{d\Sigma}{dt} = -2(\Gamma + \gamma)\Sigma + \frac{4\gamma}{\theta}\left[\langle \Delta_1 \Delta_2 \rangle + \Gamma_p \tau_b \Delta^{(2)}\right]$$

$$\frac{d}{dt}\langle \Delta_1 \Delta_2 \rangle = -4\left(\frac{2\gamma}{\theta} + \frac{1}{2\tau'}\right)\langle \Delta_1 \Delta_2 \rangle - \frac{8\gamma}{\theta}\Gamma_p \tau_b \Delta + \frac{2\Delta_0}{\tau'}\Delta^{(2)} \quad . \tag{23}$$

$$\dot{\Delta}^{(2)} = -\frac{4\gamma}{\theta}\left(\Delta^{(2)} + \Gamma_p \tau_b\right) - 2\gamma\Sigma - \frac{1}{\tau'}\left(\Delta^{(2)} - \Delta_0\right)$$

From the stationary Equation (23), using expressions (9) and (21) we obtain the stationary $\Delta^{(2)}$:

$$\Delta^{(2)} = \frac{\Gamma_p \tau_a (1 - 2\gamma \tau_b)}{1 + \Gamma_p \tau_a + 2\gamma \tau_a (1 + Sr)}, \quad Sr = \frac{\gamma}{\Gamma + \gamma} \frac{\Gamma_p \tau_a + \Gamma_p \tau_b (1 + \Gamma_p \tau_a)}{1 + \tau_a (\Gamma_p + 2\gamma)}. \tag{24}$$

Note, that necessary condition for relevance of the pump scheme shown in Figure 1 is $-1 < \Delta^{(2)} < 1$ so we have to apply restriction

$$\gamma \tau_b < 1. \tag{25}$$

The reason of restriction (25) is that we did not take into account the Coulomb blockade for reservoirs of emitter's decay, supposing large number of empty electron states in reservoirs.

Solving the third one of Equation (23) with $\Sigma \equiv 0$ we find the stationary $\Delta^{(1)}$ - population inversion of the single emitter near the nanorod. Then we observe, that if we drop the term $Sr$ in the first of Equation (24), we obtain $\Delta^{(1)}$. Thus, the term $Sr > 0$ in Equation (24) describes the contribution of plasmonic SR.

In the case of slow relaxation rate from the low state $|b\rangle$ of emitter or too high enhancement emitter radiation by nanorod, so that $1/2 < \gamma \tau_b < 1$, we do not have population inversion at $|a\rangle \rightarrow |b\rangle$ transition: according to Equation (24) $\Delta^{(2)} < \Delta^{(1)} < 0$ ($\Delta^{(1)}$ is given by Equation (24) with $Sr = 0$). Then, as one concludes from Equation (17), RQE $R < 1$. Thus the stationary population inversion has to be provided for increase RQE by SR. At the absence of population inversion in our scheme, SR reduces RQE. This result is quite similar with numerical results of.[11], [12]

We see from Equation (24) that $Sr > 0$, so if $\Delta^{(2)} > 0$ then the superradiance increases RQE given by Equation (17): as larger as bigger is $Sr$. One can see from Equation (24) that $Sr$ grows with the pump rate $\Gamma_p$.

**Figure 2a** shows normalized number $4\kappa\tau_a n^{(N)}$ of plasmons generated per emitter and found from Equation (24), (15), versus dimensionless pump rate $\Gamma_p\tau_a$ for two emitters near the nanorod (solid curves) and for single emitter near the nanorod (dashed curves) for various $\gamma\tau_a$, for

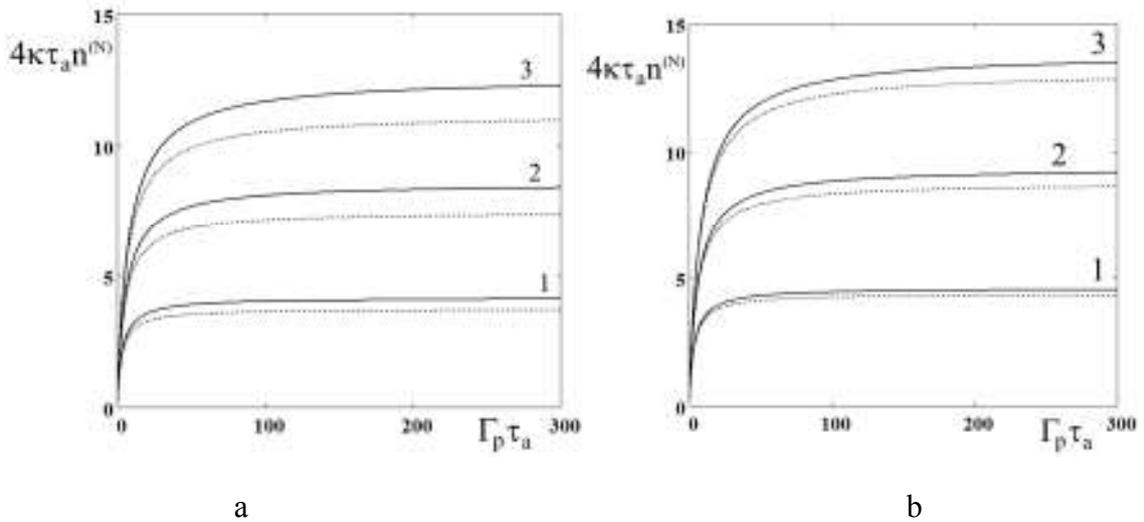

a                      b

**Figure 2.** Normalized number of plasmons $4\kappa\tau_a n^{(N)}$ generated per emitter versus dimensionless pump rate $\Gamma_p\tau_a$ for two emitters, $N = 2$ (solid curves) and for single emitter, $N = 1$, (dashed curves), for $\tau_b/\tau_a = 0.1$ and (a): low dephasing, $\Gamma_{deph}\tau_a = 0$; $\gamma\tau_b = 1$ (curves 1), $\gamma\tau_b = 1.9$ (curves 2), $\gamma\tau_b = 2.8$ (curves 3). (b) high dephasing $\Gamma_{deph}\tau_a = 10$; $\gamma\tau_b = 1.1$ (curves 1), $\gamma\tau_b = 2.2$ (curves 2), $\gamma\tau_b = 3.3$ (curves 3). Curves 2 in (a) and (b) correspond to $\gamma = \gamma_{opt}$ for the maximum increase of RQE.

$\tau_b/\tau_a = 0.1$ and negligibly small dephasing $\Gamma_{deph}\tau_a \ll 1$. **Figure 2b** shows the same for finite dephasing rate $\Gamma_{deph}\tau_a = 10$; values of other parameters are indicated in the figure capture. At high pump rate $\Gamma_p\tau_a > 10 \gg 1$ one can see the increase of RQE and the dephasing reduces this increase.

**Figure 3a,b** show RQE given by Equation (17) for the same values of parameters as for Figure 2a.b. One can conclude from Figure 3, that there is an *optimal value* of the enhanced spontaneous emission rate $\gamma = \gamma_{opt}$, corresponding to the maximum increase of RQE, and therefore the maximum efficiency of plasmonic superradiance. For $\gamma < \gamma_{opt}$ or for $\gamma > \gamma_{opt}$ RQE is smaller, than for $\gamma = \gamma_{opt}$. Because of $\gamma$ depends on the position of emitter near the nanorod, in particular, on the distance between the nanorod and the emitter, one can determine optimal positions of emitters correspondent to the maximum RQE. We'll find $\gamma_{opt}$ from Equation (24) and (17).

Because of $Sr$ grows with pump, the maximum of RQE can be obtained at high pump limit $\Gamma_p\tau_{a,b} \gg 1$. Using results (24) one can show that at $\Gamma_p\tau_{a,b} \gg 1$ RQE given by Equation (17) came to

$$R(x) = \frac{b+2x}{b+x+x^2}, \tag{26}$$

where $b = 1 + 2\Gamma_{deph}\tau_b + \tau_b/\tau_a$, $x = 2\gamma\tau_b$. The function $R(x)$ has a maximum $R = R_{max}$ at

$$x_{opt} \equiv 2\gamma_{opt}\tau_b = \frac{b}{2}\left(-1 + \sqrt{1+2/b}\right). \tag{27}$$

Absolute maximum $R = 1.155$ is reached for rapid depopulation of low levels of emitters: $\tau_b/\tau_a \to 0$ and for negligibly small dephasing $\Gamma_{deph}\tau_a \ll 1$.

Equation (27) lets us calculate $\gamma_{opt}$ for given values of other parameters. For Figure 2,3 $2\gamma_{opt}\tau_b = 0.37$ for $\Gamma_{deph}\tau_a \to 0$ and $2\gamma_{opt}\tau_b = 0.44$ for $\Gamma_{deph}\tau_a = 10$; another values of $\gamma$ on Figure 2 and 3 are taken $0.5\gamma_{opt}$ and $1.5\gamma_{opt}$. For $\Gamma_{deph} > 0$, $\gamma_{opt}$ is greater than for $\Gamma_{deph} = 0$, therefore the 999emitter-nanorod coupling is larger (see the definition of coupling coefficient Equation (13)), this is why the number of generated plasmons in Figure 2b with dephasing is larger than in Figure 2a without dephasing.

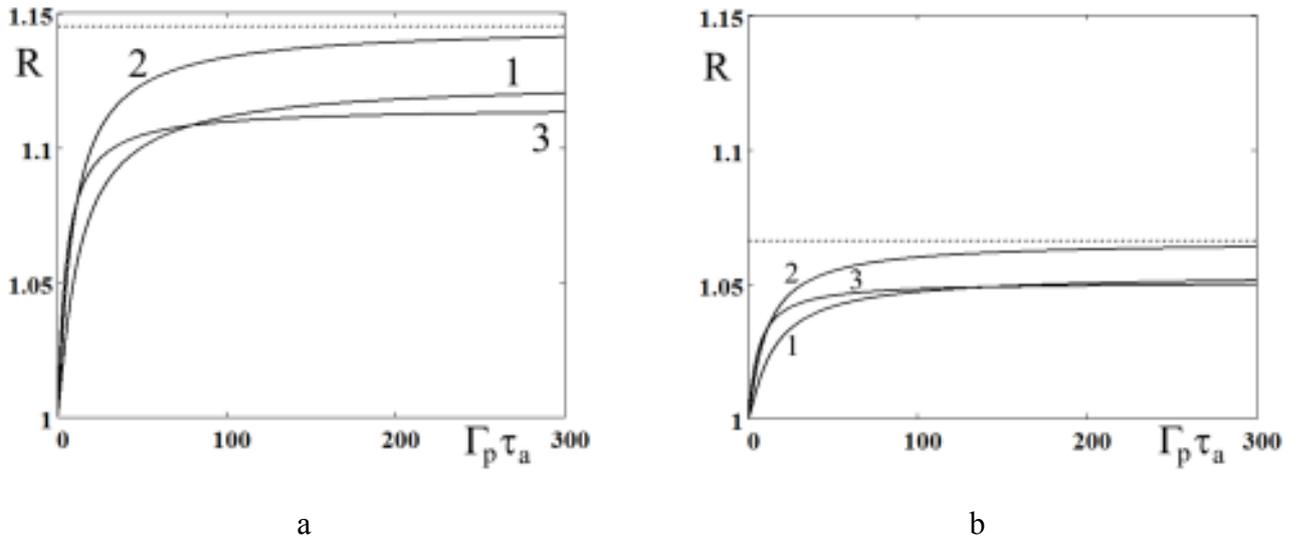

**Figure 3.** Relative RQE (a) for no dephasing, $\Gamma_{deph}\tau_a = 0$, $\gamma = 0.5\gamma_{opt}$ (curve 1); $\gamma = \gamma_{opt} = 0.37\tau_b^{-1}$ (2) and $\gamma = 1.5\gamma_{opt}$ (3); (b) for $\Gamma_{deph}\tau_a = 10$, $\tau_b/\tau_a = 0.1$, $\gamma = 0.5\gamma_{opt}$ (curve 1); $\gamma = \gamma_{opt} = 0.44\tau_b^{-1}$ (2) and $\gamma = 1.5\gamma_{opt}$ (3). Dashed red lines mark asymptotic values of $R = 1.145$ without dephasing (a) and $R = 1.07$ with dephasing (b).

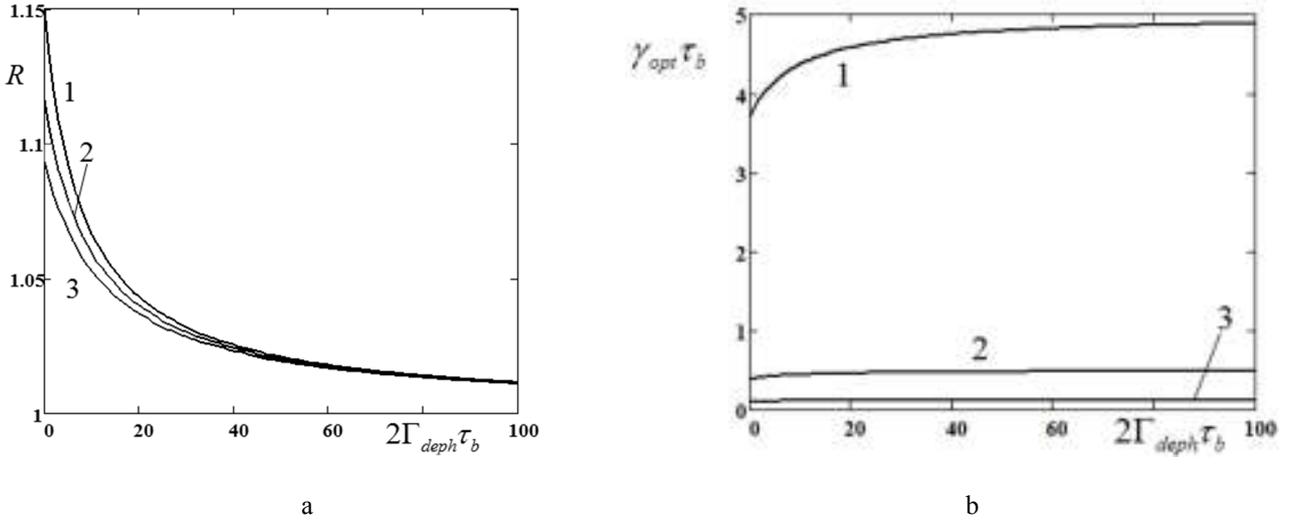

a                                          b

**Figure 4.** (a) Maximum RQE $R$ and (b) optimum of normalized enhanced spontaneous emission rate $\gamma_{opt}\tau_b$ of emitters as functions of normalized emitter dephasing rate $2\Gamma_{deph}\tau_b$ for $\tau_b/\tau_a = 0.05$ (curves 1), $\tau_b/\tau_a = 0.5$ (curves 2) and $\tau_b/\tau_a = 1$ (curves 3).

**Figure 4a** shows the maximum RQE $R_{max}$, Figure 4b – optimal value $\gamma_{opt}\tau_a = x_{opt}\tau_a/2\tau_b$ correspondent to $R_{max}$ as a functions of dimensionless dephasing rate $2\Gamma_{deph}\tau_b$ for various $\tau_b/\tau_a$.

## 5. Discussion of results

We see that positive values of $Sr$ in the denominator of the first of Equation (24) reduces the stationary population inversion per emitter $\Delta^{(2)}$ for two emitters respectively to $\Delta^{(1)}$ - for single emitter near the nanorod. For $\Delta^{(2)} > 0$, relation $\Delta^{(2)} < \Delta^{(1)}$ means increase RQE for two emitters respectively to single emitter, see Equation (17). The last stationary Equation (23) and (18) show, that $Sr$ comes from the contribution of the interference term $2\gamma\Sigma = 2\gamma(\langle\sigma_1^+\sigma_2\rangle + \langle\sigma_2^+\sigma_1\rangle)$, which is the rate of coherent interaction of emitters with each other.

Physically RQE increases because of electric field of SPP, induced by emitters, synchronies chaotic oscillations of emitter's dipole. Similar self-synchronization of oscillations of emitter's dipoles occurred through photonic modes of free space at superradiance of photons.

Noticeable synchronization takes place, when the pump of emitters is high, the pump rate substantially exceeds depopulation rates of emitter's states: $\Gamma_p \gg \tau_{a,b}^{-1}$: then the SPP electric field near the nanorod is strong and provide high emitter-emitter interaction. Oscillations of SPP field, nanorod dipole momentum and dipole momenta of emitters are noisy due to dephasing of dipoles and high absorption in the metal nanorod. However, for high pump, these noisy oscillations became *correlated* with each other. One can see (also from Equation (14) with $\gamma_1 = \gamma_2 = \gamma$) that the square of normalized dipole momentum of two emitters is

$$\sim n_1^a + n_2^a + \Sigma, \qquad (28)$$

where $n_1^a + n_2^a$ is the sum of squares of dipole momenta of the first and the second emitter, while $\Sigma \equiv \sigma_1^+\sigma_2 + \sigma_2^+\sigma_1$ is the interference term. From the stationary solution of Equation (24) we see, that $\Sigma > 0$: there is always *constructive* interference between oscillations of dipole moment of emitters at population inversion. However here we do not know what is the sign of $\Sigma$ and what is the kind of interferences in more general case, with $\gamma_1 \neq \gamma_2$, this may be a subject of future studies.

Equation (17), (24) show that the necessary condition of increase of RQE in our scheme is the population inversion at emitter transitions: $\Delta^{(2)} > 0$. If there is no population inversion, $\Delta^{(2)} < 0$ then RQE is reduced by SR. According to Equation (9) the pump scheme in Figure 1 provides population inversion at the absence of radiation: $\Delta_0 > 0$ for any pump. It is interesting to verify whether the requirement of population inversion for increasing RQE by SR preserved for other pump-relaxation schemes. Condition $\Delta^{(2)} > 0$ needed for RQE increase reminds lasing condition. However, in a difference with lasing, there is no threshold condition $\Delta^{(2)} > \Delta_{th}^{(2)} > 0$ of dominating stimulated emission, because of only spontaneous emission of plasmons (though enhanced by collective effects) takes place here: because of $\kappa\tau' \gg 1$ stationary number of plasmons $n^{(N)} \ll 1$, see Equation (15).

The maximum RQE $R = R_{max}$, given by Equation (26), is obtained at high pump $\Gamma_p \gg \tau_{a,b}^{-1}$: $R_{max}$ about 10-15% can be approached at $\Gamma_p \tau_{a,b} \geq 100$ at low dephasing rate, see Figure 3. Dephasing reduces $R_{max}$, degree of reduction can be estimated from Figure 4a. Maximum value of $R_{max}$ is reached at $\tau_a \gg \tau_b$. However for $\tau_a \gg \tau_b$ one may need too high pump rate $\Gamma_p$ to reach $R_{max}$, so it may be better to stay with largest $\tau_a \geq \tau_b$, when $R_{max}$ is already close to 15%, see Figure 3a. $R_{max}$ is reached at certain value of spontaneous emission rate $\gamma = \gamma_{opt}$ of emitter into SPP. Dependence of $\gamma_{opt}$ on parameters is given by Equation (27) and shown in Figure 4b. The most important dependence is $\gamma_{opt} \sim \tau_b^{-1}$.

10-15% of increase of RQE without dephasing seems not so high, but it is much higher than predicted degree of acceleration of spontaneous emission from two emitters by SR in free space with no dephasing and with only radiative decay: see, for example, Figure 5 from[3] or Figure 3 from.[16] Plasmonic SR is more intensive than SR in free space because of high density of states of photons near the nanorod. Note that few percent's of increase of radiative decay of two emitters (superconductive q-bits) was observed in the prove-of-principle experiment[10] in THz region, which is, as far as we aware, the only experiment on SR of two emitters in cavity. Thus, two or few emitters near metal nanostructure may be good system for experimental observation of SR in the visible or near IR spectral regions. RQE will be increased furtherly with growing the number of emitters symmetrically placed near the nanostructure.

Direct measurement of RQE is, may be, not a best way to detect and study SR near nanoparticle, because of difficulties in setting proper positions of emitters, emitter's inhomogeneity varied from sample to sample etc. However, because of the coupling of emitters through SPP provides synchronization in emitter's dipole transitions, one can expect synchronization in emitter's *blinking* at their emission into free space. Experimental manifestation of such correlated blinking may be good prove of plasmonic SR. One can expect that the maximum synchronization of blinking will take place in the same regions of parameters where we found the maximum of RQE enhancement. The next step in theoretical analysis will be direct calculations of correlations in radiation of emitters near metal nanorod.

## 6. Conclusion

We carried out fully quantum mechanical analysis of the stationary radiation from two emitters near metal nanorod. The only approximation applied was the adiabatic elimination of variables of nanorod's localized surface plasmon-polariton mode (SPP), which is good for low-Q SPP weakly coupled with emitters. Non-radiative damping and dephasing on emitter's transitions, absorption of SPP in metal nanoparticle, continuoues incoherent pump of emitters have been taken into account. We found stationary number of generated plasmons and relative quantum efficiency of generation of plasmons per emitter (RQE) for two emitters, respectively to single emitter near metal nanoparticle. We see, that RQE>1 at the population inversion in emitter's transitions. The maximum RQE is about 15% for high pump rate, low dephasing and fast relaxation from the low energy state of emitter. Maximum value of RQE is reached for certain optimal value of coupling of emitter with SPP.

The results and approach of this model help in the design of future experiments on observation of SR near metal nanostructures and make a basis for more detailed modeling of plasmonic SR. Observation of plasmonic SR of several emitters near metal nanorod may be good "proof of principle" experiment on quantum SR in dissipative environment. One can register correlations in blinking of emitters or increase in the efficiency of generation of light per emitter. Possible applications of plasmonic SR can be for excitation (also by injection current) and control of non-radiative modes in plasmonic waveguides;[13] for fast and efficient single photon sources (SPS might not be the case); for lowing threshold of plasmonic nanolasers.

7. Appendix

**7.1 Quadrupole charge density oscillations along nanorod.**

Let us describe some particular SPP in metal nanorod. We consider quadrupole oscillations of the nanorod charge density. In order to describe them we separate the nanorode into two equal parts 1 and 2 as shown in **Figure 5**.

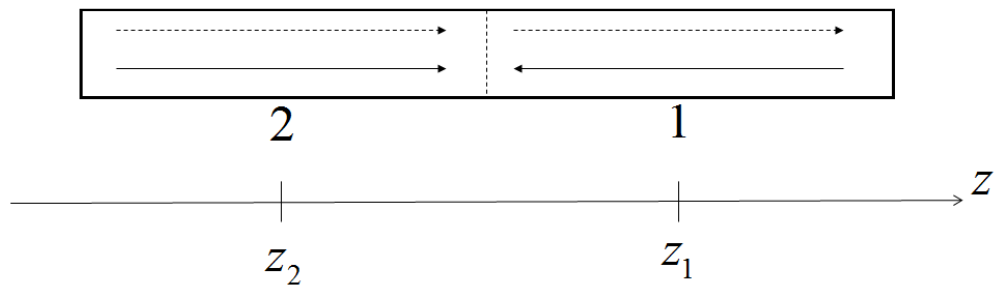

**Figure 5.** Direction of "bright" (dipole) oscillations is shown by the dashed arrows, for "dark" (quadrupole) oscillations – by solid arrows.

We describe harmonics oscillations of the charge density in the part 1 by Bose-operator $a_1$ and in the part 2 - by Bose-operator $a_2$. We describe electromagnetic field outside the nanorod as a superposition of fields from dipoles 1 and 2 located, correspondently, in points $z_1$ and $z_2$.

Charge density oscillations interact with each other through electromagnetic field inside the nanorod with the rate $\Omega$ and decay due to radiative and non-radiative losses of energy with the rate $\Gamma$. We assume strong coupling of charge density oscillations: $\Omega \gg \Gamma$ so we have two different charge oscillation eigenmodes, and we can find them neglecting by dissipation. Dissipation then will be added in each eigenmode.

Equations of motion for $a_1$ and $a_2$ (without relaxation terms) are

$$\begin{aligned} i\dot{a}_1 &= \omega a_1 - \Omega a_2 \\ i\dot{a}_2 &= \omega a_2 - \Omega a_1 \end{aligned} \qquad (A1)$$

Here $\omega$ is frequency of harmonic oscillations of only the left (the right) part of the nanorod, considered separately, without another part; $\hbar\Omega$ is the energy of the interaction between two parts. Solution of Equation (A1) is

$$\begin{aligned} a_1 &= \frac{1}{\sqrt{2}}\left[a_b e^{-i\omega_b t} + a_d e^{-i\omega_d t}\right] \\ a_2 &= \frac{1}{\sqrt{2}}\left[a_b e^{-i\omega_b t} - a_d e^{-i\omega_d t}\right] \end{aligned}, \qquad (A2)$$

where Bose-operator $a_b$ corresponds to "bright" oscillation mode: when this mode is excited dipoles 1 and 2 oscillates in phase with frequency $\omega_b = \omega - \Omega$. These are dipole oscillations of whole electron density of the nanorod. Bose-operator $a_d$ corresponds to "dark" mode: when it is excited, dipoles 1 and 2 oscillates in opposite phases with frequency $\omega_d = \omega + \Omega$. "Bright" and "dark" modes are well separated in frequencies at strong coupling, when $\Omega \gg \Gamma$. Dark mode has small radiative losses: radiation from dipoles 1 and 2 is almost canceled due to interference in the far field region: cancelling is not complete only because of finite distance $z_2 - z_1 \neq 0$ between dipoles 1 and 2. When the dark mode is excited, total dipole momentum of the nanorod is zero. Thus, the dark mode can be interpreted as a quadrupole electric mode of oscillations of the nanorod charge density.

**7.2 Coupling coefficients and Hamiltonian.**

We consider $N$ identical emitters as, for example, two emitters in points $1e$ and $2e$ near the nanorod – Figure 6.

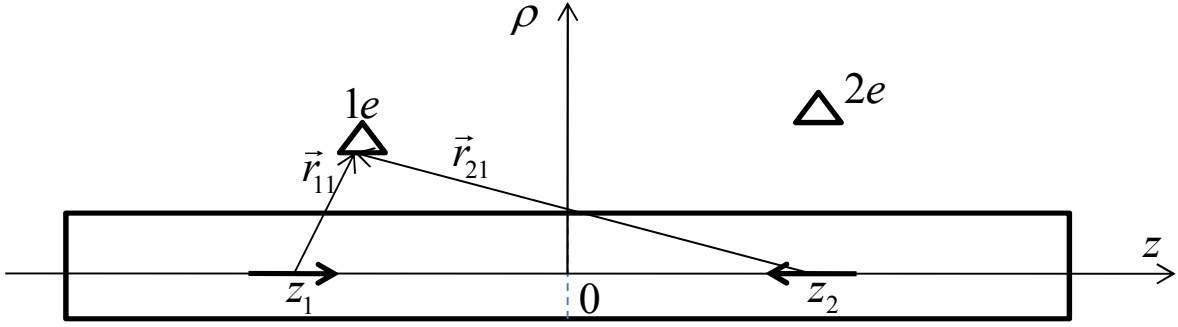

**Figure 6** Two emitters near nanorod

Let us specify the basic states of emitters. We suppose that the nanorod has axial symmetry relatively to axes $z$ in Figure 6, we have to consider only two excited states of each emitter. Transition from the ground state to one of them corresponds to the transition dipole momentum directed along axes $z$ in Figure 6, to another state – dipole momentum along axes $\rho$, perpendicular to axes $z$ in the plane made by axes $z$ and two vectors $\vec{r}_{ij}$ to that emitter – as shown in Figure 6. We approximate SPP field as the sum of fields from two dipoles in points $z_1$ and $z_2$ with dipole momentums directed along the nanorod axes $z$. Here we suppose that the emitter-emitter interaction is much weaker than the emitter-nanorod interaction so we neglect by the emitter-emitter interaction. As far as the emitter-emitter interaction is negligibly small, the axial symmetry of the nanorod lets us to consider transitions only from two emitter excited states mentioned above. Emitter transitions from the third excited state with the dipole momentum correspondent to the axes perpendicular to axes $\rho$ and $z$, do not interact with SPP.

Hamiltonian, describing the interaction of the nanorod and emitters through the electric field of SPP is

$$H = \hbar\omega_b a_b^+ a_b + \hbar\omega_d a_d^+ a_d + \hbar\omega_e \sum_{i,\alpha=z,\rho} n_{i\alpha}^a - \sum_{i,j} \hat{\vec{d}}_{ei} \hat{\vec{E}}_{ji} + \sum_{\beta=b,d} \hat{\Gamma}_\beta + \sum_{i=1}^{N} \hat{\Gamma}_{ei}. \quad (A3)$$

Here first two terms is the energy of free oscillations of bright and dark oscillating modes of electron density of the nanorod; the third term is the energy of emitters with $\omega_e$ - transition frequency of an emitter (equal for each emitter), $n_{i\alpha}^a$ - operator of the population of the upper $|a\rangle$ energy state of i-th emitter, transition from this state to ground state corresponds to $\alpha = z, \rho$ component of transition dipole momentum; fourth term describes the interaction of i-th emitter with transition dipole momentum operator $\hat{\vec{d}}_{ei}$, with electric dipole radiation field from the nanorod, described by operator $\hat{\vec{E}}_{ji}$ and originated from j-th part of oscillations of nanorod electron density:

$$\hat{\vec{E}}_{ji} = \frac{1}{2} \sum_{\omega=\omega_b, \omega_d} \left[ \vec{E}_{ji}(\vec{r}_{ji}, \omega) a_i(\omega) e^{-i\omega t} + \vec{E}_{ji}^*(\vec{r}_{ji}, \omega) a_i^+(\omega) e^{i\omega t} \right], \tag{A4}$$

where $\vec{E}_{ji}(\vec{r}_{ji}, \omega_i)$ is a c-number vector specified below, and $a_i(\omega)$ - component of the operator $a_i$, given by Equation (A2), oscillating at frequency $\omega$. According to Equation (A2) $a_1(\omega_b) = 2^{-1/2} a_b$, $a_1(\omega_d) = 2^{-1/2} a_d$, $a_2(\omega_b) = 2^{-1/2} a_b$ and $a_2(\omega_d) = -2^{-1/2} a_d$.

We suppose, for example, that emitter transition frequency $\omega_e$ is close to $\omega_d$ - the frequency of oscillations of the dark mode and far from $\omega_b$ - the frequency of oscillations of bright mode, $\vec{r}_{ij}$ is shown in Figure 6. So we'll consider the excitation of only dark SPP and neglect by the excitation of bright SPP. In this case

$$\begin{aligned}\hat{\vec{E}}_{1i} &= \frac{1}{2\sqrt{2}} \left[ \vec{E}_{1i}(\vec{r}_{1i}) a_d e^{-i\omega_d t} + \vec{E}_{1i}^*(\vec{r}_{1i}) a_d^+ e^{i\omega_d t} \right] \\ \hat{\vec{E}}_{2i} &= \frac{-1}{2\sqrt{2}} \left[ \vec{E}_{2i}(\vec{r}_{2i}) a_d e^{-i\omega_d t} + \vec{E}_{2i}^*(\vec{r}_{2i}) a_d^+ e^{i\omega_d t} \right]\end{aligned}, \tag{A5}$$

where c-number vector

$$\vec{E}_{ji}(\vec{r}_{ji}) = \frac{nd\omega_d^3}{c^3} \left[ \vec{e}_z f_z(kr_{ji}) + \vec{e}_{ji} (\vec{e}_{ji} \cdot \vec{e}_z) f_r(kr_{ji}) \right] e^{ikr_{ji}} \tag{A6}$$

Here $n$ is the refractive index of the environment of the nano-rode; $d$ is the matrix element of dipole momentum of oscillations of the half of the nano-rode; $c$ is speed of light in vacuum; $\vec{e}_z$ is unit vector along z-axes in Figure 6; $k = n\omega_d/c$ is the wave number, $r_{ji} = |\vec{r}_{ji}|$; unit vector $\vec{e}_{ji} = \vec{r}_{ji}/r_{ji}$;

$$f_z(x) = x^{-1} + ix^{-2} - x^{-3}, \qquad f_r(x) = -x^{-1} - 3ix^{-2} + 3x^{-3}, \qquad x \equiv kr_{ij}. \qquad (A7)$$

Component $\alpha = z, \rho$ of dipole momentum operator of emitter is

$$\hat{d}_{i\alpha} = d_e \left( \sigma_{i\alpha} e^{-i\omega_d t} + \sigma_{i\alpha}^+ e^{i\omega_d t} \right). \qquad (A8)$$

Here $d_e$ is the matrix element of transition of emitter; operator $\sigma_{i\alpha}$ - for transition from the upper state to the lower state of i-th emitter with transition dipole momentum oriented along axes $\alpha = z, \rho$.

Because of the bright mode is not excited, we can re-write Hamiltonian (A3) only with the dark SPP. We use rotating wave approximation:

$$H_{RWA} = \sum_{i,\alpha=z,\rho} \left\{ -\hbar\delta n_{i\alpha}^a - \frac{d_e}{2\sqrt{2}} \left[ \left( \vec{E}_{1i} - \vec{E}_{2i} \right)_\alpha \sigma_{i\alpha}^+ a_d + c.c \right] \right\} + \hat{\Gamma}_d + \sum_{i=1}^N \hat{\Gamma}_{ei} \qquad (A9)$$

Here $\vec{E}_{ij} \equiv \vec{E}_{ij}(\vec{r}_{ij})$ given by Equation (A6); $\left( \vec{E}_{1i} - \vec{E}_{2i} \right)_\alpha$ means projection of the vector $\vec{E}_{1i} - \vec{E}_{2i}$ to axes $\alpha = z, \rho$; detuning $\delta = \omega_d - \omega_e$.

Hamiltonian (A9) and further treatment can be simplified taking into account that each emitter interacts with circular polarized electric field. Indeed, operator $\hat{\vec{E}}_i$ of electric field of dark SPP in the position of i-th emitter is:

$$\hat{\vec{E}}_i = \frac{1}{2\sqrt{2}} \sum_{\alpha=\rho,z} \left[ \left| \left( \vec{E}_i \right)_\alpha \right| \vec{e}_\alpha a_d e^{-i\omega_d t + i\varphi_{i\alpha}} + c.c. \right]. \qquad (A10)$$

Here $\vec{E}_i = \vec{E}_{1i} - \vec{E}_{2i}$, $(\vec{E}_i)_\alpha$ is complex number, $(\vec{E}_i)_\alpha = |(\vec{E}_i)_\alpha| e^{i\varphi_{i\alpha}}$, $\vec{e}_\alpha$ is unit vector in the direction of axes $\alpha = z, \rho$. We can re-write Equation (A10) as

$$\hat{\vec{E}}_i = \frac{1}{2\sqrt{2}} |\vec{E}_i| \left[ \left( \cos\theta_i \vec{e}_z e^{i\varphi_{iz}} + \sin\theta_i \vec{e}_\rho e^{i\varphi_{i\rho}} \right) a_d e^{-i\omega_d t} + c.c. \right] \tag{A11}$$

where $|\vec{E}_i| = \sqrt{|(\vec{E}_i)_z|^2 + |(\vec{E}_i)_\rho|^2}$, $\cos\theta_i = |(\vec{E}_i)_z|/|\vec{E}_i|$. Electric field given by Equation (A11) is in the superposition of two states: with the polarization directed along $\vec{e}_z$ and along $\vec{e}_\rho$. This field interacts with a superposition of excited states of i-th emitter: $|z\rangle_i$ and $|\rho\rangle_i$ - with transition dipole moment parallel to axes z, and $\rho$, correspondently. This superposition is

$$|a\rangle_i = \left( \cos\theta_i e^{i\varphi_{iz}} |z\rangle_i + \sin\theta_i e^{i\varphi_{i\rho}} |\rho\rangle_i \right) e^{-i\omega_d t}. \tag{A12}$$

In principle, one can write the quantum state $|a\rangle$ of the field, which is similar superposition as (A12), only instead of emitter states $|z\rangle_i$ and $|\rho\rangle_i$ there are field states $|z\rangle$ and $|\rho\rangle$ with polarizations directed along $\vec{e}_z$ and $\vec{e}_\rho$, respectively.

Introducing $\sigma_i$ - transition operator from the state $|a\rangle_i$ to ground state $|0\rangle_i$ of i-th emitter and $n_i^a$ - operator of the population of this state we can write instead of Hamiltonian (A9)

$$H_{RWA} = \hbar \sum_{i=1}^{N} \left[ -\delta n_i^a - \Omega_i \left( \sigma_i^+ a_d + c.c. \right) \right] + \hat{\Gamma}_d + \sum_{i=1}^{N} \hat{\Gamma}_{ei}, \tag{A13}$$

where the coupling coefficient $\Omega_i = \dfrac{d_e |\vec{E}_i|}{2\sqrt{2}\hbar}$ (Rabi frequency) is real c-number. Thus, we can describe each emitter as a two-level system. Hamiltonian (A13) with $a_d \equiv a$ and $\delta \equiv \omega - \omega_e$ is used in Section 2 as Hamiltonian (1) for the derivation of equations of motion.

**References**

[1] R. H. Dicke, Phys. Rev. **1954**, *93*, 99.

[2] L. I. Men'shikov, Phys. Usp. **1999**, *42*, 107.

[3] S. Haroche, M. Gross, *Phys. Rep.* **1982**, *93*, 301.

[4] L. Allen, J. Eberly, *Optical Resonance and Two-Level Atoms,* Dover Publications, **1987**.

[5] T. Brandes, *Phys. Rep.* **2005**, *408*, 315.

[6] V. V. Temnov, U. Woggon, *Opt. Expr.* **2009**, *17*, 5774.

[7] V. V. Temnov, U. Woggon, *Phys. Rev. Lett.* **2005**, *95*, 243602.

[8] A. Auffèves, D. Gerace, S. Portolan, A. Drezet, M. F. Santos, *New Journ. Phys.* **2011**, *13*, 093020.

[9] A. Moelbjerg, P. Kaer, M. Lorke, B. Tromborg, J. Mørk, *IEEE Journ. Quant. Electron.* **2013**, *49*, 945.

[10] J. A. Mlynek, A. A. Abdumalikov, C. Eichler, A. Wallraff, *Nature Comm.* DOI: 10.1038/ncomms6186.

[11] F. Jahnke, C. Gies, M. A mann, M. Bayer, H.A.M. Leymann, A. Foerster, J. Wiersig, C. Schneider, M.Kamp, H. Sven, *Nature Comm.* DOI: 10.1038/ncomms11540.

[12] E. Mascarenhas, D. Gerace, M. F. Santos, A. Auffeves, *Phys.Rev.A* **2013**, *88*, 063825.

[13] S. Zhang, H. Xu, *Small,* **2014**, *10*, 4264.

[14] V. Weisskopf, E.Wigner, *Zeitschrift für Physik,* **1930***,* 63, 54.

[15] P. W. Milonni, P. L. Knight, *Phys. Rev. A* **1974***, 10,* 1096.

[16] I. E. Protsenko, *JETP* **2006***, 103,* 167.

[17] J. Barthes, A. Bouhelier, A. Dereux, G. Colas des Francs, *Scient. Rep.* **2013***,* 1.

[18] I. E. Protsenko, A. V. Uskov, V. M. Rudoy, *JETP* **2014***, 119,* 265.

[19] V. N. Pustovit, T. V. Shahbazyan, *Phys. Rev. B* **2010***, 82,* 075429.

[20] I. E. Protsenko, A. V. Uskov, *Quantum Electron.* **2015***, 45,* 561.

[21] M. Moskovits, *Rev. Mod. Phys.* **1985**, 57, 783.

[22] I. E. Protsenko, A. V. Uskov, **2016***, http://arxiv.org/abs/1604.08302*.

[23] M. Scheibner, T. Schmidt, L. Worschech, A. Forchel, G. Bacher, T. Passow, D. Hommel, *Nature Phys.* **2007** *3,* 106.

[24] T. Berstermann, T. Auer, H. Kurtze, M. Schwab, D. R.Yakovlev, M. Bayer, J. Wiersig, C. Gies, F. Jahnke, *Phys. Rev. B* **2007***, 76,* 165318.

[25] R. G. DeVoe, R. G. Brewer, *Phys. Rev. Lett.* **1996***, 76,* 2049.

[26] Y. Zhang R. Zhang, Q. Wang, Z. Zhang, H. Zhu, J. Liu, S. Lin, Y. B. Pun Edwin, *Optics Express* **2010***, 18,* 4316.

[27] Y. Rostovtsev, S. S. Dhayal, *Phys. Rev. A* **2016***, 93,* 043405.

[28] A. Delga, J. Feist, J. Bravo-Abad, F. J. Garcia-Vidal, *Phys. Rev. Lett.* **2014***, 112,* 253601.

[29] S. Zhang, L. Chen, Y. Huang, H. Xu, *Nanoscale* **2013**, *5*, 6985.

[30] I. E. Protsenko, M. Travagnin, *Phys. Rev. A* **2001**, *65*, 013801.

[31] I. Protsenko, P. Domokos, V. Lefe`vre-Seguin, J. Hare, J. M. Raimond, L. Davidovich, *Phys. Rev. A* **1999**, 59, 1667.